\documentclass[article,preprint,amsmath,amssymb,aps,floatfix,superscriptaddress,longbibliography]{revtex4-2}

\usepackage{graphicx}
\usepackage{dcolumn}
\usepackage{bm}
\usepackage[usenames,dvipsnames]{xcolor}
\usepackage[mathlines]{lineno}% Enable numbering of text and display math
\usepackage{tikz}

\newcommand{\perSA}[1]{{\textcolor{orange}{XXX for Stefano: #1}}}
\newcommand{\rv}[1]{{\textcolor{black}{#1}}} % rv stands for revised
\newcommand{\LC}[1]{\textcolor{blue}{XXX Comment LC: #1 XXX}}

\newcommand{\SI}{Supplementary}
\newcommand{\MM}{Methods}
\newcommand{\um}{~\mu \mathrm{m}}
\newcommand*\circled[1]{\tikz[baseline=(char.base)]{
            \node[shape=circle,draw,inner sep=2pt] (char) {#1};}}

\begin{document}

%\preprint{APS/123-QED}

\title{A unified state diagram for the yielding transition of soft colloids}% Force line breaks with \\

\author{Stefano Aime$^*$}
 \affiliation{Laboratoire Charles Coulomb (L2C), Universit\'e Montpellier, CNRS, Montpellier, France}
 \affiliation{Present address: ESPCI, Paris, France}
 \email{stefano.aime@espci.fr}
\author{Domenico Truzzolillo}
 \affiliation{Laboratoire Charles Coulomb (L2C), Universit\'e Montpellier, CNRS, Montpellier, France}
\author{David J. Pine}
\affiliation{New York University}
\author{Laurence Ramos}
\affiliation{Laboratoire Charles Coulomb (L2C), Universit\'e Montpellier, CNRS, Montpellier, France}
\author{Luca Cipelletti$^*$}
\affiliation{Laboratoire Charles Coulomb (L2C), Universit\'e Montpellier, CNRS, Montpellier, France}
\affiliation{Institut Universitaire de France}
 \email{luca.cipelletti@umontpellier.fr}

\date{\today}

\begin{abstract}

%\SA{Currently: we are about 3100+ words long, without abstract fig captions etc. Nat Materials and Nat. Physics (both no cost): 2000-3000 words, Nature comm (4530E): 5000 words max. PNAS: 6 pages, but can be exceeded (if you pay...). I think we are about 7-8 PNAS pages. In all Nature-something journals, the abstract is typically 150 words; it contains a brief account of the background and rationale of the work, followed by a statement of the main conclusions introduced by the phrase "Here we show" or some equivalent. The abstract below is about 158 words long.}
%\newline

Concentrated colloidal suspensions and emulsions are amorphous soft solids, widespread in technological and industrial applications and studied as model systems in physics and material sciences. \rv{They} are easily fluidized by applying a mechanical stress, \rv{undergoing a yielding} transition \rv{that still lacks a} unified description. Here, we investigate yielding in three classes of \rv{repulsive} soft solids, using analytical and numerical modelling and experiments probing the microscopic dynamics and mechanical response under oscillatory shear. We find that at the microscopic level yielding consists in a transition between two distinct dynamical states\rv{, which we rationalize by proposing} a lattice model with dynamical coupling between neighboring sites, leading to a unified state diagram for yielding. \rv{Leveraging the analogy with Wan der Waals’s phase diagram for real gases, we show that distance from a critical point} plays a major role in the emergence of first-order-like \textit{vs} second-order-like features in yielding, thereby reconciling previously contrasting observations on the nature of the transition.
\end{abstract}

%\pacs{Valid PACS appear here}
\maketitle

%\SA{An introduction (without heading) of up to 500 words of referenced text expands on the background of the work (some overlap with the summary is acceptable), followed by a concise, focused account of the findings, ending with one or two short paragraphs of discussion.}

The yielding transition of soft glassy systems is of great relevance both in technological and industrial applications and at a fundamental level~\cite{bonn_yield_2017}. Despite profound differences in their microscopic structural features, yielding occurs with very similar macroscopic features in systems as diverse as colloidal and nanoparticle suspensions~\cite{koumakis_complex_2013%gibaud_shear_induced_2009,chen_microscopic_2020
}, emulsions~\cite{mason_yielding_1996,knowlton_microscopic_2014,rogers_microscopic_2018}, star polymers~\cite{rogers_oscillatory_2011%,christopoulou_ageing_2009
} and microgels~\cite{ketz_rheology_1988%carrier_nonlinear_2009,cloitre_yielding_nodate
}. This suggests the presence of an underlying general framework, which has been addressed in recent experimental, theoretical and numerical works~\cite{sollich_rheology_1997,seth_micromechanical_2011}, leading to contrasting results. \rv{Measurements of the macroscopic viscoelastic properties suggest that yielding develops progressively} as the system is driven far from the linear regime~\cite{koumakis_complex_2013,mason_yielding_1996,rogers_oscillatory_2011,%christopoulou_ageing_2009,
ketz_rheology_1988,%cloitre_yielding_nodate,
donley_elucidating_2020}. Various models such as the soft glassy rheology~\cite{sollich_rheology_1997}, the mode coupling theory~\cite{brader_nonlinear_2010,voigtmann_2014}
and \rv{fluidity~\cite{picard_simple_2002,benzi_unified_2019,liu_mean-field_2018}
%fluidity~\cite{picard_simple_2002,carrier_nonlinear_2009}
or on-lattice~\cite{sainudiin_microscopic_2015}} models reproduce the evolution of viscoelastic parameters across yielding. %However, these models typically rely on a number of phenomenological parameters that account for nonlinear rheolog,liuMeanFieldScenarioAthermal2018y, with little or no insight into the underlying microscopic processes.
\rv{Recent experiments and simulations probing microscopic quantities indicate that yielding is associated with an increase of particle mobility ~\cite{keim_yielding_2013,fiocco_oscillatory_2013,knowlton_microscopic_2014,hima_nagamanasa_experimental_2014,jeanneret_geometrically_2014,kawasaki_macroscopic_2016,leishangthem_yielding_2017,rogers_microscopic_2018,edera_deformation_2021}, suggesting that it may be described as a dynamic transition between a quiescent, solid-like state and a dynamically active, fluid-like state, bearing analogies with equilibrium phase transitions, an approach similar to that used in the past to describe other flow-induced transitions~\cite{dusek_shear-induced_2009}. Note, however, that this description does not take into account the ultra-slow relaxations that typically occur in soft solids even at rest~\cite{l._cipelletti_universal_2003,madsen_beyond_2010}. These works suggested contrasting scenarios for the yielding transition.} Some systems exhibit features typical of a first-order transition, such as a discontinuous jump of the \rv{particle mobility~\cite{jeanneret_geometrically_2014,knowlton_microscopic_2014,kawasaki_macroscopic_2016,leishangthem_yielding_2017,rogers_microscopic_2018} or of structural symmetries~\cite{denisov_sharp_2015}}, the coexistence of dynamically distinct states~\cite{jeanneret_geometrically_2014}, and hysteresis~\cite{divoux_rheological_2013}. By contrast, \rv{in other cases} yielding is described as a rather continuous transition~\cite{bocquet_kinetic_2009,hima_nagamanasa_experimental_2014}, with features such as sluggish dynamics~\cite{keim_yielding_2013,fiocco_oscillatory_2013,knowlton_microscopic_2014,hima_nagamanasa_experimental_2014}, \rv{enhanced fluctuations~\cite{knowlton_microscopic_2014,nordstrom_dynamical_2011} and growing length scales~\cite{hima_nagamanasa_experimental_2014}} typically associated with a second-order transition. Thus, the nature of the yielding transition remains elusive: there is a dearth of experiments and modelling addressing the mechanical response and the microscopic dynamics of a class of soft materials sufficiently diverse to allow for a general description of yielding.

\begin{figure}[ht]
\includegraphics[width=\columnwidth]{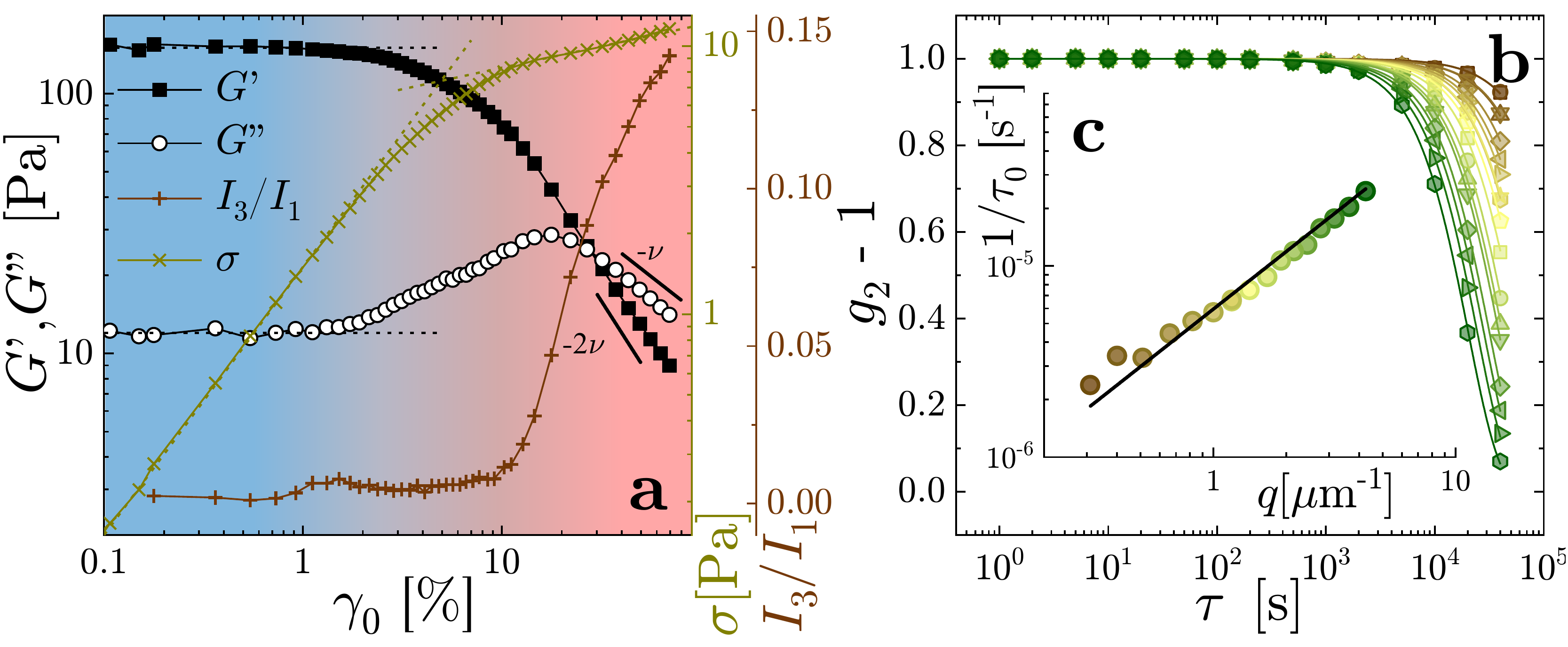}
\caption{\textbf{Viscoelasticity and spontaneous dynamics of a dense microgel suspension.} a): Oscillatory rheology for a microgel suspension at effective volume fraction $\varphi=1.5$ (sample \rv{M40s}). Left axis: first-harmonic storage ($G'$, black squares) and loss ($G''$, white circles) moduli \textit{vs} strain amplitude $\gamma_0$, at fixed $\omega = 0.157~\mathrm{rad~s}^{-1}$. Right axis: first-harmonic stress amplitude ($\sigma$, gold crosses) and normalized third harmonic component of the stress signal ($I_3/I_1$, brown pluses). Rheological quantities evolve smoothly from solid-like behavior (small $\gamma_0$, blue shades) to fluid-like response (red shades). b): Correlation functions measured for the same system at rest, for scattering vectors $0.2~\mu m^{-1} \le q \le  4.5~\mu m^{-1}$, increasing from brown to green shades. The correlation functions are fitted with compressed exponential functions, with compression exponent \rv{$\beta_0=1.65>1$} (lines). c): Relaxation rate \rv{$1/\tau_0$} \textit{vs} scattering vector. Same color code as in b). The line is a fit with slope 1, indicating ballistic dynamics. %\perSA{y axis of c): replace $\Gamma_s$ by $1/\tau_0$}
\label{fig:rheology}}
\end{figure}

Here, we establish a unified view of the yielding transition of \rv{repulsive} soft colloids by combining experiments probing both microscopic and macroscopic quantities with theoretical modelling and numerical simulations. We investigate samples of three kinds: concentrated suspensions of microgel particles (M) and charged silica nanoparticles (N), and dense emulsions (E) (see~\MM~for details).
All samples exhibit qualitatively similar behavior in oscillatory shear tests at frequency $\omega$ and at variable strain amplitude $\gamma_0$, as exemplified by Fig.~\ref{fig:rheology}a for microgels. For small enough $\gamma_0$, $G'$ and $G''$, the storage and loss moduli, are independent of $\gamma_0$, $G'>>G''$ and the stress amplitude $\sigma$ grows linearly with $\gamma_0$, indicative of a predominantly elastic, linear response. As $\gamma_0$ is increased, a gradual transition to the nonlinear regime is observed: $G'$ and $G''$ deviate from their low-$\gamma_0$ behavior, with $G''$ going through a maximum and eventually exceeding $G'$. Deviations from a purely harmonic response become non-negligible, as shown by the growth of the normalized third harmonic amplitude $I_3/I_1$ of the stress response. At the largest strain amplitudes, $\sigma$ grows sublinearly with $\gamma_0$ and both moduli decay as power laws: $G'\propto \gamma_0^{-2\nu}$ and $G''\propto \gamma_0^{-\nu}$, with a sample-dependent exponent $\nu$ in the range 0.6-0.75, see~\SI~Table SI1.

%--FFFFFFFFFFFFFFFFFFFFFFFFFFFFFFFFFFFFFF
%--FFFFFFFFFFFFFFFFFFFFFFFFFFFFFFFFFFFFFFFFF

The range of strain amplitudes over which rheological quantities signal the transition from solid-like to fluid-like behavior is quite broad, making it difficult to determine the nature of the yielding transition~\cite{%christopoulou_ageing_2009,
knowlton_microscopic_2014,donley_elucidating_2020}. To gain a deeper insight on yielding, we couple rheometry to measurements of the microscopic dynamics, using dynamic light scattering or differential dynamic microscopy (see~\MM). Both methods quantify the dynamics via the intensity correlation function $g_2(\tau)-1$, which decays from 1 to 0 as microscopic displacements grow beyond a length scale $\pi/q \approx (0.1-1)\um$ set by the scattering vector $q$.%~\cite{berne76,giavazzi_digital_2014}.

\begin{figure}[ht]
\includegraphics[width=\columnwidth]{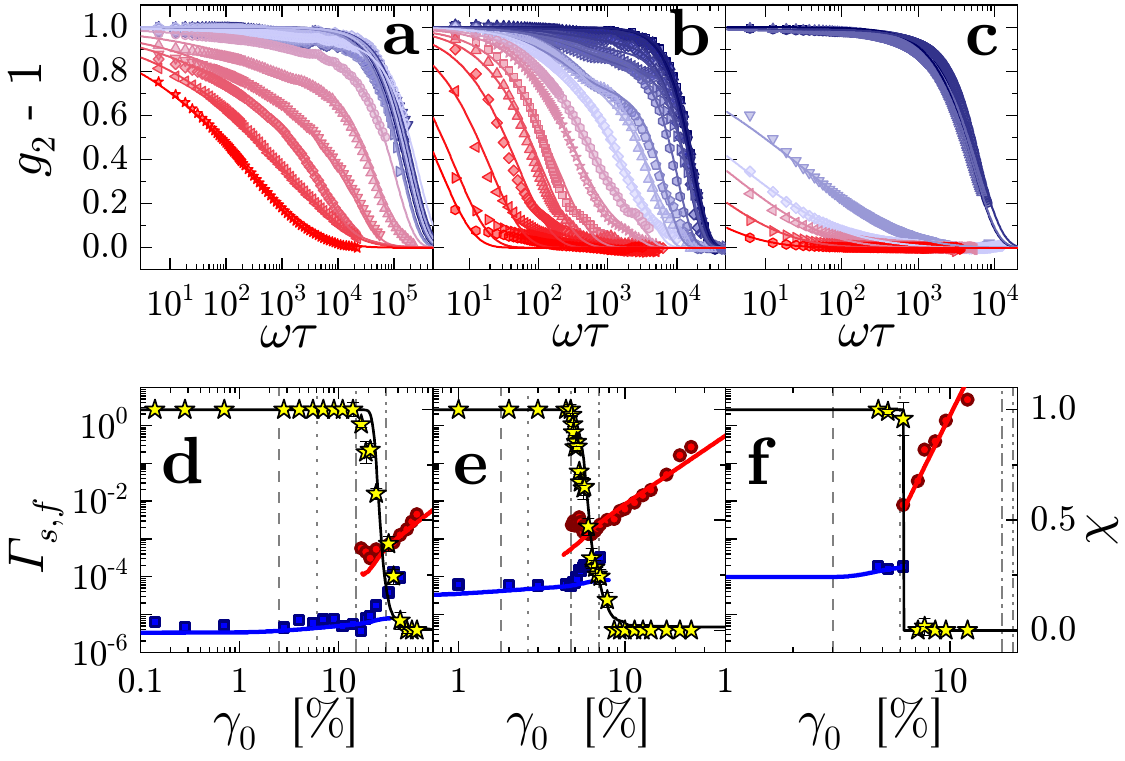}
\caption{\textbf{Yielding of soft solids as a dynamic transition}. a)-c): Intensity correlation functions under oscillatory shear, for concentrated microgels (a, \rv{M2s}), nanoparticles (b, N45\%), and emulsions (c, E74\%), plotted \textit{vs} the normalized time delay $\omega \tau$, with $2\pi/\omega$ the period of the oscillations. \rv{$\omega = \pi~\mathrm{rad/s}$ (resp., $2\pi~\mathrm{rad/s}$) for samples M2s and N45\% (resp., E74\%)}. The strain amplitude $\gamma_0$ increases from blue to red shades, spanning the rheological yielding transition (see~\MM~for sample details and~\SI~Table SI1 for the probed deformation ranges). Symbols: experimental data. Lines: fits using Eq.~\ref{eqn:g1_dblexp}. The compressed exponent $\beta_s$ of the slow mode is \rv{1.3, 1.9, 1.4 for microgels, nanoparticles, and emulsions, respectively.} The stretching exponent $\beta_f$ of the fast mode is a fit parameter shared between all data for a given sample, yielding $\beta_f=0.4$, 0.9, and 0.3 for \rv{M2s, N45\% and E74\%}, respectively. The fitting parameters for the same samples are shown in d)-f). Left axes: $\gamma_0$ dependence of the normalized rates \rv{$\Gamma_s$} (slow mode, blue squares) and \rv{$\Gamma_f$} (fast mode, red circles). Right axes: relative amplitude $\chi$ of the slow mode  (stars). Lines: numerical results of the general model for the yielding transition discussed in the text, see~\SI~Table \rv{SI2} for details on the parameter values. \rv{In d)-f), the vertical lines indicate, from left to right, the onset of the increase of $G"$, the onset of deviations from linearity of $\sigma(\gamma_0)$, the maximum of $G"$, the crossover between $G'$ and $G"$}.
\label{fig:corrfunc}}
%(dashed) (dotted)  (dash-dotted) (dash-dot-dotted)
\end{figure}

The spontaneous dynamics measured at rest are similar for all samples, and are well described by a slow, compressed exponential relaxation: \rv{$g_2^{(s)}(\tau)-1=\exp\left[-2\left(\tau/\tau_0\right)^{\beta_0}\right]$},
Fig.\ref{fig:rheology}b, with sample-dependent values of the spontaneous relaxation \rv{rate $1/\tau_0$ and of the exponent $\beta_0>1$}, see~\SI~Table SI1.
These dynamics are ballistic, as indicated, for the microgels, by the linear dependence of \rv{$1/\tau_0$} with $q$, Fig.\ref{fig:rheology}c. Similar spontaneous dynamics have been reported for many other jammed or glassy soft samples at rest, and are attributed to the slow relaxation of quenched internal stresses~\cite{l._cipelletti_universal_2003}. To investigate the microscopic dynamics under deformation, we apply an oscillatory shear with angular frequency $\omega$ and measure $g_2-1$ stroboscopically, for $\tau$ values a multiple of the oscillation period $2 \pi/\omega$. The dynamics probed by this echo protocol~\cite{hebraud_yielding_1997%,petekidis_rearrangements_2002
} are only sensitive to irreversible rearrangements, either spontaneous or induced by shear.
Figures~\ref{fig:corrfunc}a-c reveal striking similarities of the overall behavior of the correlation functions across all samples. Under small strain amplitudes, the dynamics are independent of $\gamma_0$, while they increasingly accelerate with growing strain at larger $\gamma_0$. Concomitantly, the shape of $g_2-1$ evolves from a steep compressed exponential decay at low $\gamma_0$ to a stretched shape at large $\gamma_0$.
Data at all strain amplitudes are very well fitted by the following expression:
\rv{
%++eeeeeeeeeeeeeeeeeeeeeeeee
\begin{equation}
	g_2(\tau)-1 =\left\{ \chi\exp\left[-\left(\Gamma_s \omega \tau\right)^{\beta_s}\right]
		+\left(1-\chi\right)\exp\left[-\left(\Gamma_f \omega \tau\right)^{\beta_f}\right] \right\}^2 \,,
	\label{eqn:g1_dblexp}
\end{equation}
%++eeeeeeeeeeeeeeeeeeeeeeeee
}
where $\beta_f$ and $\beta_s$ are (sample-dependent) constants, whereas $\chi$, $\Gamma_s$ and $\Gamma_f$ vary with $\gamma_0$. \rv{The dimensionless relaxation rates for the slow and fast relaxation modes, normalized by the oscillation frequency $\omega$, are designated by $\Gamma_s$ and $\Gamma_f$, respectively.}

Figures~\ref{fig:corrfunc}d-f show the strain dependence of the normalized relaxation rates, \rv{$\Gamma_{s,f}$}, and of the slow mode amplitude $\chi$, obtained by fitting Eq.~\ref{eqn:g1_dblexp} to the correlation functions of Figs.~\ref{fig:corrfunc}a-c. Three regimes can be distinguished.
For small enough strain amplitudes, $g_2-1$ relaxes through a single, slow compressed exponential mode ($\chi=1$), with a stretching exponent $\beta_s \ge1$ (see~\SI~Table SI1) and a strain-independent relaxation rate \rv{$\Gamma_s$ close to that at rest, $1/\omega\tau_0$}. For the microgels, oscillatory tests at $\omega = 0.157 ~\mathrm{rad~s}^{-1}$ and $\omega= 3.14~\mathrm{rad~s}^{-1}$ indicate no dependence of \rv{the slow mode with $\omega$ (in physical units)}, further confirming that the dynamics observed in this regime are unaffected by shear and simply correspond to the sample spontaneous relaxation.
As $\gamma_0$ exceeds a threshold value, correlation functions become strain-dependent. A second, faster mode, characterized by a stretched exponential relaxation, adds to the spontaneous relaxation mode, whose relative amplitude $\chi$ rapidly decays from 1 to 0 with increasing $\gamma_0$. Finally, for sufficiently large $\gamma_0$, $\chi \approx 0$: the correlation functions are well fitted by a single stretched exponential relaxation, with \rv{$\Gamma_f$} increasing as $\gamma_0^n$, with a sample-dependent exponent \rv{$1<n<8$} (red symbols in Figs~\ref{fig:corrfunc}d-f). In this regime, we check for microgels that \rv{the fast relaxation rate, in physical units, is proportional to}  $\omega$, as expected in the case of dynamics fully dominated by rearrangements induced by strain oscillations. Moreover, we find that for the microgels and emulsions \rv{$\Gamma_f$} scales as $q^2$ (see~\SI~Figs. \rv{SI8-SI9}), the hallmark of diffusive motion, as also reported recently \rv{in simulations~\cite{kawasaki_macroscopic_2016} and in experiments on} other kinds of microgels under large shear strain~\cite{edera_deformation_2021}. This shear-induced diffusive behavior at large $\gamma_0$ is analogous to the dynamics of \textit{equilibrated} dense colloidal suspensions at rest~\cite{vanmegen_measurement_1998a,weeks_threedimensional_2000a}, and contrasts with the ballistic behavior at small strain or at rest. Our experiments thus show that at the microscopic level yielding corresponds to a \rv{transition between ultraslow, ballistic relaxations at small $\gamma_0$ (unaccounted for in previous works) and fast, diffusive relaxations beyond yielding. In analogy to the recently reported abrupt change of microscopic quantities such as the particle mean squared displacement or diffusivity~\cite{knowlton_microscopic_2014,kawasaki_macroscopic_2016,edera_deformation_2021}, the amount of irreversible rearrangements~\cite{rogers_microscopic_2018,keim_yielding_2013,jeanneret_geometrically_2014}, and the size of avalanches~\cite{leishangthem_yielding_2017}, the correlation functions measured in our experiments exhibit a marked change in a narrow interval of $\gamma_0$, indicative of a transition sharper than for rheological quantities, compare the stars and the vertical lines in Figs.~\ref{fig:corrfunc}d-f}.

%++FFFFFFFFFFFFFFFFFFFFFFFFFFFFFFFFFFFFFF
%++FFFFFFFFFFFFFFFFFFFFFFFFFFFFFFFFFFFFFF

%Our light scattering setups allow for simultaneous rheometry and microscopic dynamics measurements and thus for a direct comparison of the rheological and dynamic yielding transitions. We find that the latter is significantly sharper: the slow mode amplitude $\chi$ typically drops from 1 to 0 over a factor of 2 in $\gamma_0$ (Figs.~\ref{fig:corrfunc}d-e), with the emulsion samples exhibiting an even sharper transition (Fig.~\ref{fig:corrfunc}f and~\SI~Fig.~\rv{SI15}). By contrast, the rheological transition spans about one decade in strain, from the onset of deviations from linear behavior of $G'$ and $G''$ to the crossover strain $\gamma_x$ where $G'' = G'$ (Fig.~\ref{fig:rheology}a). For samples M and N, the onset of the dynamic transition occurs at a strain close to the onset of anharmonic stress response, while the vanishing of the slow mode approximately coincides with $\gamma_x$ (see the gray shaded region in Fig.~\ref{fig:corrfunc}d,e).
%Our experiments thus show that at the microscopic level yielding corresponds to a \rv{transition between two well-distinct dynamical states}.

To rationalize these findings, we introduce a simple model. The sample is coarse-grained on a lattice; each lattice site is attributed a relaxation rate $\Gamma_i$ \rv{that depends on both the spontaneous relaxation at rest $1/\omega \tau_0$ and shear-induced rearrangements $\Gamma_{sh,i}$:}
%++eeeeeeeeeeeeeeeeeeeeeeeee
\begin{subequations}\label{eq:gamma_sh}
\begin{align}
\rv{\Gamma_{i}}& \rv{=1/\omega \tau_0+\Gamma_{sh,i}}
  \label{eq:rate_addtivity}\\
   \rv{\Gamma_{sh,i}} & \rv{= \frac{K}{\gamma_0^{-n}+ N^{-1}\sum_j \frac{\alpha_{ij}}{\Gamma_i\Gamma_j}}\,,}
  \label{eq:shear_corentibution}
\end{align}
\end{subequations}
%++eeeeeeeeeeeeeeeeeeeeeeeee
\rv{where $K$ is a constant whose physical meaning will be discussed later, and where the sum in the r.h.s. of Eq. \ref{eq:shear_corentibution} runs over the $N$ nearest neighbors of site $i$, with $\alpha_{i,j}$ coupling constants between the dynamics of sites $i$ and $j$. Equation~\ref{eq:rate_addtivity} states that the overall relaxation rate is the sum of two independent contributions: $1/\omega \tau_0$, the spontaneous relaxation rate, and $\Gamma_{sh,i}(\gamma_0)$, the site- and strain amplitude-dependent rate of the additional relaxation induced by shear. A similar additive rule has been invoked in mode coupling-based models~\cite{derec_rheology_2001,
%wyss_strain-rate_2007,
miyazaki_nonlinear_2006,hess_yielding_2011}, which postulated $\Gamma_{sh,i} \propto \gamma_0^n$. However, this form of the shear-induced relaxation rate yields a smooth growth of $\Gamma_i$ with $\gamma_0$, rather than a well-defined transition. Instead, we propose in Eq.~\ref{eq:shear_corentibution} an alternative ansatz for the shear-induced relaxation rate. It is the simplest expression that accounts for the following physical ingredients:  i) $\Gamma_{sh,i}$ should vanish for small strain amplitudes, implying $\Gamma_i \approx 1/\omega \tau_0$ in the $\gamma_0 \rightarrow 0$ limit; ii) in the opposite limit of large $\gamma_0$, the dynamics should be dominated by the externally imposed shear, implying $\Gamma_i \rightarrow K\gamma_{0}^n$, as measured in our experiments for the fast mode; iii) in the intermediate regime, the shear-induced dynamics should be ruled not only by the external drive, but also by the interactions between neighboring sites, which we expect to slow down the system relaxation, as modelled by the sum term in the r.h.s. of Eq.~\ref{eq:shear_corentibution}. The latter is chosen in the spirit of dynamic facilitation models for the spontaneous relaxation of glassy systems~\cite{biroli_perspective_2013}, where sites with a higher-than-average relaxation rate  facilitate the relaxation of neighboring sites.}

We \rv{start by considering the mean-field version of the model}, where the coupling constants and thus the relaxation rates are taken to be identical for all sites, \rv{$\alpha_{ij}\equiv \alpha$ and $\Gamma_i \equiv \Gamma$}. The mean field model can be solved analytically by recasting Eqs.~\ref{eq:gamma_sh} as
%+eeeeeeeeeeeeeeeeeeeeeeeee
\begin{equation}
\rv{	\left(\Gamma-\frac{1}{\omega \tau_0}\right)\left(\gamma_0^{-n}+\frac{\alpha}{\Gamma^2}\right) = K \,,}
	\label{eqn:eos}
\end{equation}
%+eeeeeeeeeeeeeeeeeeeeeeeee
with $\alpha$ an average coupling constant. This equation is formally identical to the Van der Waals (VdW) equation of state ruling the vapor-liquid transition of real gases, with pressure $p$ volume $V$ and temperature $T$ in the VdW's equation replaced by $\gamma_0^{-n}$, \rv{$\Gamma$}, and $K$, respectively. The spontaneous non-dimensional rate \rv{$1/\omega \tau_0$} and the coupling constant $\alpha$ play the role of the molecular volume and molecular interaction parameter in VdW's law, respectively.

\begin{figure}[ht]
\includegraphics[width=\columnwidth]{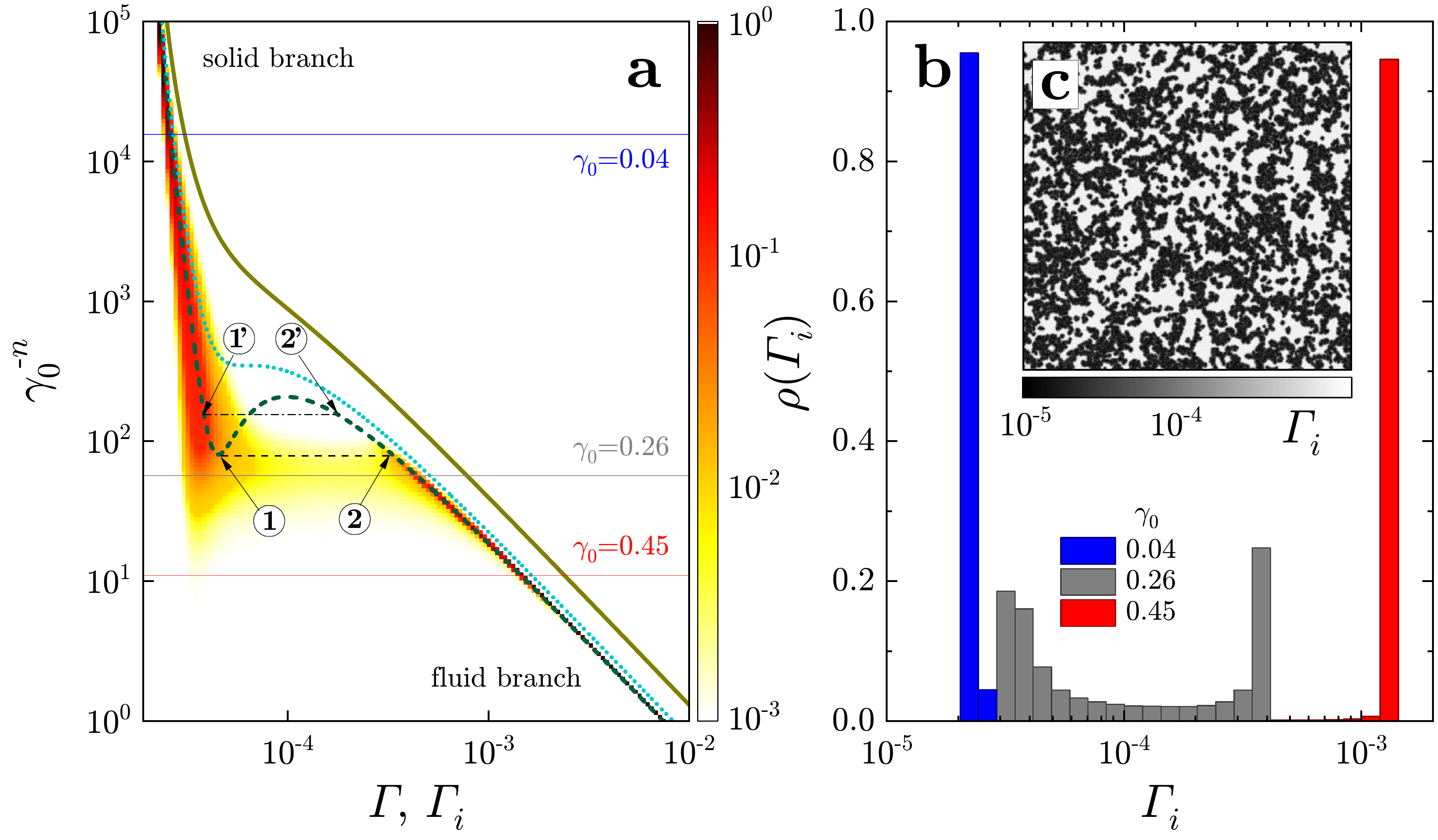}
%{FIG3_v6_lowres.pdf}
%XXX TO DO: go back to FIG3-v6.pdf in the final version!
\caption{\textbf{Theoretical and numerical modelling of the yielding transition.}
a) Lines: mean field model, Eq.~\ref{eqn:eos}, for three $K$ values: \rv{$K=4~10^{-3}$} (sub-critical, dashed line), \rv{$K=K_c=4.2~10^{-3}$ (critical iso-$K$}, dotted line) and \rv{$K=5.9~10^{-3}$} (super-critical, continuous line). \rv{The circled numbers, arrows and horizontal lines indicate the yielding transition, see text for details.} Color shades and color bar: probability distribution of the local relaxation rates \rv{$\Gamma_i$} in simulations of the model with \rv{disorder and parameters chosen to reproduce the data for the microgels M2s shown in Fig.~\ref{fig:corrfunc}d:
%OLD VALUES, should have been changed by Stefano
%$K=1.25~10^{-3}$, \rv{$1/\tau_0=1.11~10^{-5}~\mathrm{s}^{-1}$}, $\bar\alpha=1.67~10^{-8}$, $n=2$, $\omega=\pi~\mathrm{rad~s}^{-1}$.
$K=4~10^{-3}$, $1/\tau_0=2~10^{-5}~\mathrm{s}^{-1}$, $\bar\alpha=9~10^{-8}$, variance of $\alpha_{i,j}=8.9~10^{-16}$, $n=3$}, $\omega=\pi~\mathrm{rad~s}^{-1}$. b): Probability distribution of the local relaxation rate for strain amplitudes corresponding to the fluid (blue), coexistence (gray) and solid-like (red) regimes, same colors as the corresponding horizontal lines in a). c): Snapshot of the simulated system corresponding to the bimodal distribution in b), showing the coexistence of regions with low (dark shades) and high (light shades) relaxation rates. In b) and c), the parameters are the same as for the model with disorder in a).}
\label{fig:simul}
\end{figure}

In experiments, the strain amplitude is the control parameter, typically plotted on the $x$ axis. In Fig.~\ref{fig:simul}a, we rather choose \rv{$\Gamma$} as the abscissa, to emphasize the analogy of `iso-$K$' solutions of Eq.~\ref{eqn:eos} with VdW isotherms in a $pV$ diagram. We find that \rv{in the mean field model} $K$ plays a key role in differentiating samples that exhibit a yield transition from samples that are predominantly fluid-like at all $\gamma_0$, as illustrated by the three curves of Fig.~\ref{fig:simul}a. For $K$ larger than a critical value $K_c$, solid line in Fig.~\ref{fig:simul}a, $\gamma_0^{-n}$ decreases monotonically with increasing \rv{$\Gamma$}. This corresponds to the smooth growth \rv{--- with no yielding transition--- of the relaxation rate of concentrated yet equilibrated colloidal fluids upon applying a mechanical drive~\cite{zausch_equilibrium_2008} }.
For $K<K_c$, by contrast, \rv{$\gamma_0^{-n}(\Gamma)$} becomes non-monotonic (dashed line in Fig.~\ref{fig:simul}a): within a finite range of strain amplitudes, a unique value of $\gamma_0$ is now associated with multiple values of \rv{$\Gamma$}.
In a VdW fluid, this feature is associated with the vapor-liquid phase transition: upon compression, the system jumps from the vapor branch to the fluid branch of the isotherm line at a pressure set by the minimization of the free energy \rv{and corresponding to Maxwell's equal area rule}. In our model, non-monotonic iso-$K$ curves are associated with yielding. Starting from a solid at rest and increasing progressively $\gamma_0$, the system descends the \textit{solid-like} (left) branch of the equation of state, corresponding to small and nearly constant \rv{$\Gamma$}. \rv{In the representation of Fig.~\ref{fig:simul}a, portions of the equation of state with positive slope are nonphysical, because they correspond to faster relaxation rates attained at lower strain amplitudes. Thus, the system has to jump from the left branch to the right, fluid branch, which constitutes yielding in our model. We find that in the mean field limit of the model, the jump occurs at the minimum of the iso-$K$ line, from point $\circled{1}$ to point $\circled{2}$ in Fig.~\ref{fig:simul}a. Introducing disorder smears the transition and, in the limit of small disorder, the yield strain is shifted to smaller values, approaching a value set by the equivalent of Maxwell's equal area rule~\cite{inpreparation} (points $\circled{1'~}$ and $\circled{2'~}$)}. Finally, the dotted line of Fig.~\ref{fig:simul}a represents the critical iso-$K$: in analogy to the VdW's critical isotherm, it has an inflection point but no local minimum. Here, it separates systems that are fluid-like at all $\gamma_0$ from systems that are solid-like at small enough $\gamma_0$.

%----FFFFFFFFFFFFFFFFFFFFFFFFFFFFFFFFFFFFFFFFF
%---FFFFFFFFFFFFFFFFFFFFFFFFFFFFFFFFFFFFFFFFFFF

The mean field model, Eq.~\ref{eqn:eos}, describes yielding as a first-order transition between two dynamically distinct states, accounting for both the linear and the fully fluidized regimes. However, it fails to properly capture the gradual onset of the fast-relaxation mode and the regime of intermediate strain amplitudes where both modes coexist. Quenched disorder is known to smear out first-order transitions~\cite{%imry_influence_1979,hui_random_field_1989,
berker_critical_1993,bellafard_effect_2015}. To explore the role of disorder in our case, we solve numerically the full model, Eq.~\ref{eq:gamma_sh},  using \rv{model parameters that fit the microgels data of Fig.~\ref{fig:corrfunc}d} (details in~\MM).
In the presence of disorder, \rv{$\Gamma_i$} varies from site to site: representative probability distributions $\rho(\rv{\Gamma_i})$ are reported for three strain amplitudes in Fig.~\ref{fig:simul}b. In agreement with experiments, three different regimes are seen: (1) under small strain amplitudes, $\rho(\rv{\Gamma_i})$ is unimodal, peaked around a small value comparable to the relaxation rate at rest; (2) under intermediate strain amplitudes, $\rho(\rv{\Gamma_i})$ becomes bimodal as a consequence of the appearance of a second, faster mode characterized by a rate $\rv{\Gamma_f}$, typically well separated from $\rv{\Gamma_s}$ and growing with $\gamma_0$; (3) under large strain amplitudes, the slow mode vanishes and $\rho(\rv{\Gamma_i})$ is again unimodal, but is now sharply peaked around $\rv{\Gamma_f}\propto \gamma_0^n$.

We associate the bimodal nature of $\rho(\rv{\Gamma_i})$ at intermediate $\gamma_0$ with the coexistence of slow and fast relaxation modes observed experimentally, which smears the transition with respect to the mean field prediction (compare the dashed line and the distribution of $\rv{\Gamma_i}$ indicated by the color shades in Fig.~\ref{fig:simul}a).
A spatial map of the local relaxation rates reveals that slow and fast relaxing sites form a coarse structure (Fig.~\ref{fig:simul}c), %, reminiscent of the disordered configuration of the coupling constants $\alpha_{ij}$ \LC{what do you mean? if the $\alpha_{ij}$ are randomly chosen, their spatial distribution should also be random, whereas in Fig. 3c one has the impression that patches with a well-defined length scale emerge}.
consistent with the spatial localization of highly mobile regions observed in the single-cycle dynamics of sheared emulsions~\cite{knowlton_microscopic_2014}. The separation between the two modes allows one to extract from $\rho(\rv{\Gamma_i})$ two well-defined values of $\rv{\Gamma_s}$ and $\rv{\Gamma_f}$, as well as the relative weight $\chi$ of the slow mode. A suitable choice of the model parameters $K$, $n$, and of \rv{a} log-normal distribution of $\alpha$ (see~\SI~Table SI2) reproduces the experimental strain dependence of $\rv{\Gamma_s}$, $\rv{\Gamma_f}$ and $\chi$  (lines in Fig.~\ref{fig:corrfunc}d-f). The good agreement between experimental data and numerical results over up to two decades in applied strain supports the model and highlights that disorder is indeed at the origin of the dynamic coexistence spanning a finite range of $\gamma_0$.

One of the most powerful consequences of VdW's theory is the law of corresponding states, predicting identical properties for distinct fluids, provided that they all have the same pressure, volume, and temperature relative to the corresponding values at the critical point. Inspired by the law of corresponding states, we re-express Eq.~\ref{eqn:eos} using reduced variables:
%++eeeeeeeeeeeeeeeeeeeeeeeeee
\begin{equation}
	\left(\rv{\Gamma_r}-\frac{1}{3}\right)\left({\gamma_r}^{-n}+\frac{3}{{\rv{\Gamma_r}}^2}\right)=\frac{8}{3}K_r \,,
	\label{eqn:isotherms_adimensional}
\end{equation}
%++eeeeeeeeeeeeeeeeeeeeeeeeeee
where $\rv{\Gamma_r=\Gamma/\Gamma_c}$, $\gamma_r=\gamma_0/\gamma_{0,c}$, $K_r = K/K_c$, and where the values of the various parameters at the critical point, designated by the subscript $c$, are given in Table~\ref{tab:critical}. For the mean-field model, the coordinates of the critical point are derived  by imposing that both the first and the second derivative of $\gamma_0(\rv{\Gamma})$ vanish, in analogy to VdW's equation of state. For the model with disorder, we use reduced variables obtained from Table~\ref{tab:critical}  with the mean-field $\alpha$ replaced by the average value $\bar\alpha$ of the site-dependent $\alpha_{i,j}$.

%++tttttttttttttttttttttttttttttt
\begin{table}[ht]
    \centering
    \begin{tabular}{|c|c|c|}
    \hline
    $\gamma_{0,c}$ & $\rv{\Gamma_c}$ & $K_c$ \\
    \hline
    \hline
    $\left( \frac{27} {\alpha \rv{\omega^2 \tau_0^2}}\right )^{1/n}$  & $\rv{\frac{3}{\omega \tau_0}}$ & $\rv{\frac{8 \alpha \omega \tau_0}{27}}$ \\
    \hline

    \end{tabular}
    \caption{Values of the strain amplitude $\gamma_0$, relaxation rate $\rv{\Gamma}$ and $K$ parameter at the critical point as predicted by the mean field model for the yielding transition. %For the model with disorder, we define the critical point as in the mean field version, replacing $\alpha$ by the average coupling constant $\bar\alpha$.
    }
    \label{tab:critical}
\end{table}
%++ttttttttttttttttttttttttttttttt

\begin{figure}[ht]
\includegraphics[width=1\columnwidth]{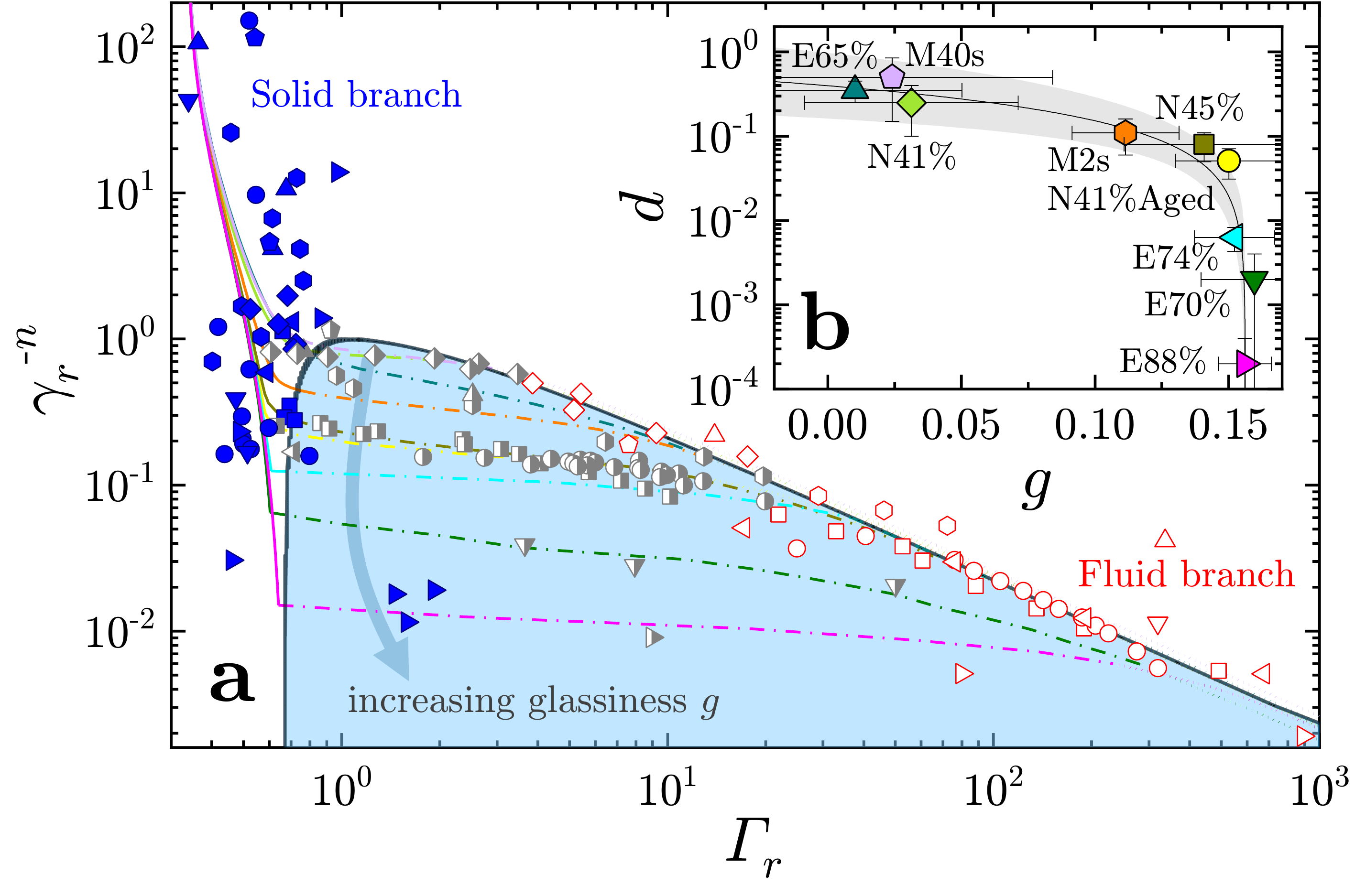}
\caption{  \textbf{Unified state diagram for the yielding transition.} a): Yielding state diagram for all samples, using reduced variables (see text). Blue, solid (resp., red, open) symbols: samples with a single slow (resp., fast) relaxation mode. Gray half-filled symbols: coexistence of the slow and fast modes. The symbol shape is the same as in b), where the samples are identified by labels, see~\MM. The continuous, dotted and dash-dotted lines indicate the solid, fluid and coexistence branches of the iso-$K$ lines obtained from simulations of the model with disorder (same color codes as in b)). Area shaded in light blue below the black line: region inaccessible to the mean field model, corresponding to the coexistence of fast and slow relaxation modes in experiments and simulations of the model with disorder. b): Disorder $d= \sigma_\alpha^2/\bar\alpha^2$ of the coupling constants $\alpha_{i,j}$ (see Eq.~\ref{eq:gamma_sh}) as a function of `glassiness', defined as the reduced distance form the critical point, $g=1-K_r$ (see also the arrow in a)).}
\label{fig:phasediagram}
\end{figure}

Figure~\ref{fig:phasediagram}a shows the unified yielding state diagram for soft colloids obtained using reduced variables. For each sample, we tune $\bar\alpha$ and the variance $\sigma^2_{\alpha}$ of the coupling constants distribution, the spontaneous relaxation rate $\rv{1/\tau_0}$ and the exponent $n$ in order to reproduce the strain dependence of $\rv{\Gamma_{s,f}}$ and $\chi$, as exemplified in Fig.~\ref{fig:corrfunc}d-f. Using these fitting parameters, we re-express the experimental variables in terms of the reduced variables of Fig.~\ref{fig:phasediagram}a. In this representation, all samples fall on nearly the same solid and fluid branches, characterized respectively by a single, compressed exponential slow mode (small $\rv{\Gamma_r}$, blue solid symbols in Fig.~\ref{fig:phasediagram}a) and a single, fast stretched exponential mode (large $\rv{\Gamma_r}$, red open symbols). This collapse is remarkable, given the diversity of the microscopic structure of the investigated samples, which in turn results in marked differences in the sensitivity to shear, compare e.g. the steep growth of $\rv{\Gamma_s}$ with applied strain for the emulsions to the gentler increase for microgels and nanoparticles (Figs.~\ref{fig:corrfunc}d-f). \rv{Furthermore, by analyzing data at various $q$ vectors for E samples, we find that the collapse is robust with respect to the choice of the probed length scale, see~\SI}. At intermediate $\rv{\Gamma_r}$, within the region inaccessible to the mean field model, a fast mode and a slow mode coexist (gray half-filled symbols in Fig.~\ref{fig:phasediagram}a), as predicted by the model with disorder; the abscissa used for these points is the weighted average of the fast and slow relaxation rates (see~\MM).

In the coexisting region, samples' properties vary markedly and systematically with $K_r$, which suggests classifying all systems according to this parameter. Since $K_r < 1$ corresponds to glassy samples and $K_r > 1$ to equilibrated fluids, we quantify `glassiness' of samples with $K_r < 1$ by $g = 1-K_r$, which increases as the iso-$K_r$ curves move downward (see arrow in Fig.~\ref{fig:phasediagram}a) away from the critical curve ($g=0$, $K_r=1$).  \rv{For samples of the same kind, the trend in $g$ (see Fig.~\ref{fig:phasediagram}b and Table SI2 in \SI) agrees with the behavior intuitively associated with a lesser or greater glassiness. For emulsions and microgels, we find that in general the higher $\varphi$ the more glassy the sample. Consistent with the notion that with age systems evolve towards deeper states in the glassy phase, we find that $g$ increases with age for sample N41\%. Finally, one expects $g$ to increase with $\omega$, because glassy samples fall increasingly out of equilibrium as the time scale of the driving becomes shorter. This is indeed what is seen in our experiments, compare samples M40s and M2s. Beyond these comparisons, the notion of glassiness paves the way for a quantitative comparison between samples of different nature (e.g. emulsions \textit{vs} microgels) or probed according to different protocols.} \rv{Keeping in mind that each iso-$K_r$ curve --and thus each $g$ value--} is characterized by a different disorder parameter of the coupling constants $d = \sigma_\alpha^2 / \bar{\alpha}^2$, we find a remarkable negative correlation between the glassiness and $d$, which, as shown in Fig.~\ref{fig:phasediagram}b, is characterized by a simple master curve.

%The notion of glassiness paves the way for a quantitative comparison between samples of different nature (e.g. emulsions \textit{vs} microgels) or probed according to different protocols (e.g. different $\omega$ or aging conditions). Within a given system, $g$ evolves as intuitively expected: for emulsions, the glassiness increases with volume fraction; for nanoparticles, aging increases $g$; for microgels, $g$ increases with $\omega$, because as the frequency increases the sample is driven at a rate increasingly faster than its spontaneous relaxation rate.

%--FFFFFFFFFFFFFFFFFFFFFFFFFFFFFFFFFFFFFFFFFF
%--FFFFFFFFFFFFFFFFFFFFFFFFFFFFFFFFFFFFFFFFFFFFFFFF

\begin{figure}[ht]
\includegraphics[width=1\columnwidth]{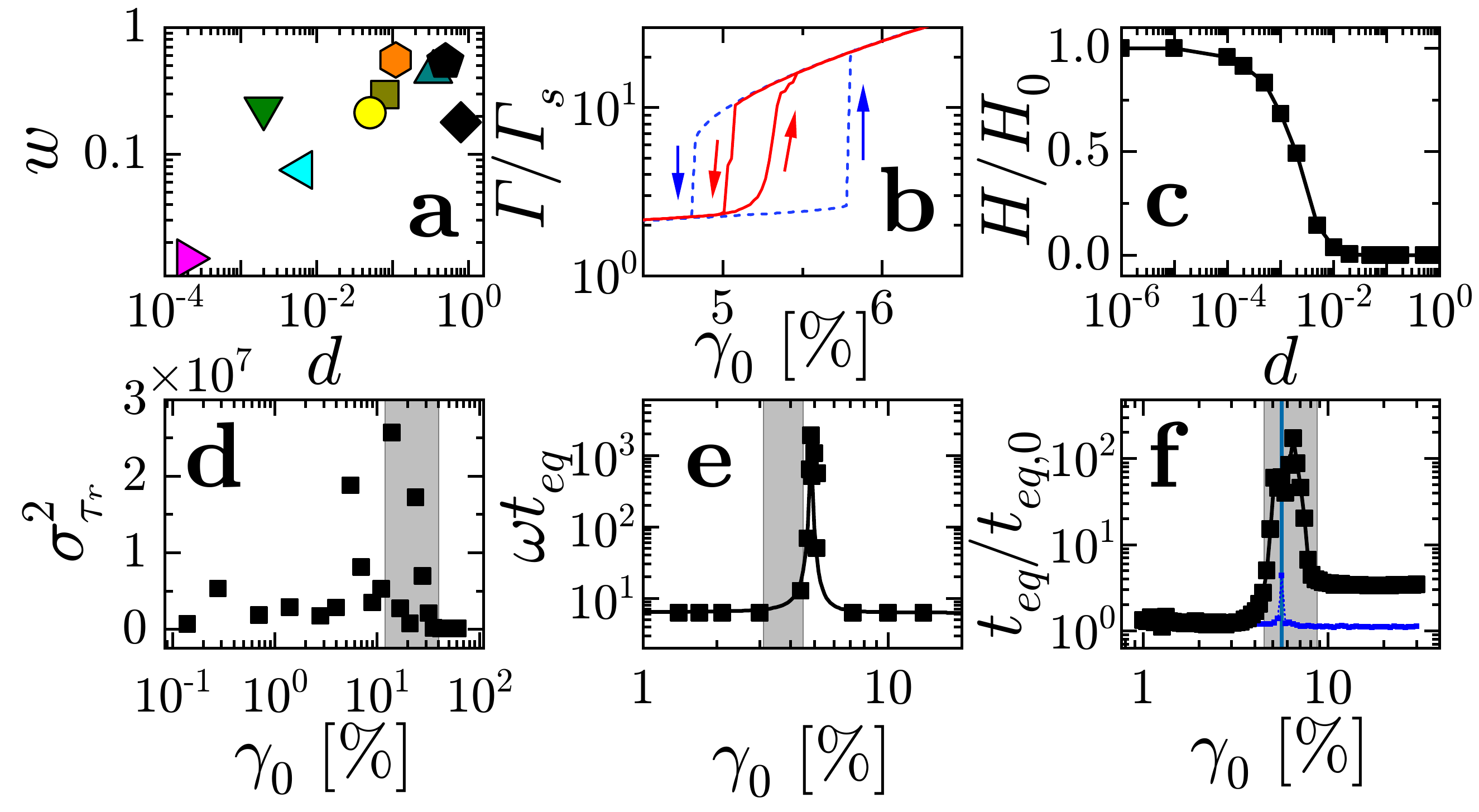}%{FIG5}
\caption{  \textbf{Distance from a critical point governs the nature of the yielding transition.} Top row: features typical of a first order transition are enhanced as glassiness increases ($d$ decreases). a): Relative width $w$ of the transition in experiments probing the microscopic dynamics across yielding (see~\MM~for details). b): Hysteresis at yielding, data from simulations with $d=0$ (dashed lines, blue shades) and  $d= 0.02$ (solid lines, red shades). c): Area $H$ of the hysteresis loop in simulations with disorder, normalized by that in the $d=0$ limit. Bottom row: features typical of a second order transition emerge close to the critical point (small $g$, large $d$). d): Fluctuations of the dynamics, quantified by the temporal variance $\sigma^2_{\tau_r}$ of the relaxation time of the intensity correlation functions, are strongly enhanced at the transition. Data for sample \rv{M2s}. e): Normalized time \rv{$t_{eq} \omega$} to attain stationary dynamics for sample N41\%.  The lines are a guide to the eye. f): Number of iterations required for numerical solutions of the model with disorder to converge. Data are normalized by their corresponding value at  $d=0$. Black symbols and line: $d= 0.1$; blue line: $d=10^{-4}$. In d)-f), the gray shaded area indicates the range of the dynamical transition (see~\MM~for details). Panels b, c, f: same mean field parameters as in Figs.~\ref{fig:corrfunc}b,e.}
\label{fig:order-transition}
\end{figure}

The correlation between glassiness and disorder has also deep implications on the nature of the yielding transition. We find that the most glassy samples, for which $d$ approaches the $d=0$ mean field limit, exhibit features typical of a first-order transition, as predicted by the mean field model (Fig.~\ref{fig:order-transition}a-c). Figure~\ref{fig:order-transition}a shows $w$, the relative width of the $\gamma_0$ range where fast and slow modes coexists in experiments, demonstrating a dramatic increase of the transition sharpness as $d$ decreases. Figures~\ref{fig:order-transition}b-c demonstrate hysteresis, another distinctive feature of first-order transitions (data from simulations). As seen in Fig.~\ref{fig:order-transition}c, hysteresis is largest for the smallest disorder and vanishes when departing from the mean field limit.

Conversely, systems close to the critical $K$ and hence with large $d$ (small $g$) exhibit features usually associated with second order transitions, illustrated in Fig.~\ref{fig:order-transition}d-f. Figure~\ref{fig:order-transition}d shows for sample \rv{M2s} that temporal fluctuations of the relaxation rate are strongly enhanced at the transition. Figures~\ref{fig:order-transition}e-f demonstrate sluggishness, another feature of second-order transitions: in both experiments (Fig.~\ref{fig:order-transition}e) and simulations (Fig.~\ref{fig:order-transition}f) on systems with large $d$ (small $g$), the time to attain a stationary state dramatically increases around the transition.

%--FFFFFFFFFFFFFFFFFFFFFFFFFFFFFFFFFFFFFFFFFF
%--FFFFFFFFFFFFFFFFFFFFFFFFFFFFFFFFFFFFFFFFFFFFFFFF

The unified yielding state diagram established here demonstrates the universal nature of the solid-fluid transition in soft glassy systems under oscillatory shear. %Our model differs from previous approaches by introducing a direct link between the microscopic dynamics and the macroscopic drive, and by including the
Our model differs from previous approaches by introducing a direct link between the macroscopic drive and the microscopic dynamics, and by including in the latter the contribution of ultra-slow, spontaneous relaxations; as a result, slow and fast relaxation modes may coexist. These two modes might be associated with the formation of shear bands organized in the shear gradient direction~\cite{divoux_shear_2016,radhakrishnan_shear_2016}. Our light scattering experiments cannot test \rv{directly} this hypothesis; however, \rv{our numerical data, microscopy experiments on emulsions, and photon correlation imaging measurements on sample N45\%} are inconsistent with shear banding (\SI~Sec. IV), \rv{rather pointing to the formation of fast-relaxing domains similar to those of Fig.~\ref{fig:simul}c}. This suggests that the coexistence of slow- and fast-relaxing regions is a generic feature, not necessarily implying the \rv{structuration} in shear bands. A key result of our work is the emergence of different features in the transition depending on the distance from the critical point. This indicates a promising way to reconcile apparently conflicting reports in previous studies of yielding.

Finally, the model proposed here does not depend on the details of the interaction between the microscopic constituents of the system: we thus expect it to provide a general framework for the yielding transition, \rv{possibly including in systems with attractive interactions~\cite{pham_multiple_2002,gibaud_heterogeneous_2010}. }

\section*{\label{sec:matmeth}\MM}

\subsection*{Samples} Table~\ref{tab:samples} summarizes the main sample and experimental parameters. PNIPAM microgels (M) were synthesized by emulsion polymerization as described in \cite{truzzolillo_overcharging_2018}, and were suspended in a 2~mM $\mathrm{NaN_3}$ aqueous solution to prevent bacterial growth. The microgel radius at $T=23~^\circ \mathrm{C}$ is 294~nm, as measured by dynamic light scattering (DLS) in a diluted suspension. Sample preparation and characterization, including the determination of the effective volume fraction $\varphi$, are described in~\cite{philippe_glass_2018}. Note that the effective volume fraction is larger than one, due to the particle softness. %Sample structure has shown to be unaffected by shear in the probed range of amplitudes and frequencies \perSA{`has shown...': by whom? how? Any ref? Only true for microgels? May be just cut...}.
Charged nanoparticle systems (N) were prepared by concentrating an acqueous suspension of silica particles (Ludox TM50, from Sigma Aldrich), as described in~\cite{philippe_glass_2018}. The particles have a hydrodynamic radius of 23~nm, as measured by DLS in the dilute limit. To improve the scattering signal, the samples were seeded with 200~nm-diameter polystyrene particles at extremely low volume fraction, $\phi_{PS}<10^{-6}$. We check that this seeding has no measurable impact on the rheological properties microscopic dynamics of the samples. Concentrated emulsions (E) were prepared by dispersing polydimethylsiloxane droplets into a water/glycerol matrix, as described in~\cite{knowlton_microscopic_2014}. The resulting droplets have an average size of $2.4\um$ and 20\% polydispersity. All samples were initialized by applying a preshear, \rv{see Sec.~IIa of the \SI~for details}. Rheology and microscopic dynamics started immediately after applying the preshear for all samples, except for sample N41\%Aged, which was left at rest for 12h prior to measurements.
\begin{table}[ht]
    \centering
    \begin{tabular}{l|c c c c l l}
         Sample ID & $\varphi$ & $T$ (s) & \rv{$\omega$ [rad/s]} & \rv{$q$ $[\um^{-1}]$} & setup & \rv{$\mathbf{q}$ orientation}\\
         \hline
         \rv{M2s} & 1.5 & 2 & \rv{3.14} & \rv{5} & PCI & \rv{vorticity} \\
         \rv{M40s} & 1.5 & 40 & \rv{0.157} & \rv{0.1-5} & SALS & \rv{vorticity} \\
         N41\% & 0.41 & 2 & \rv{3.14} & \rv{30} & PCI & \rv{shear gradient}\\
         N41\%Aged & 0.41 & 2 & \rv{3.14} & \rv{30} & PCI & \rv{shear gradient}\\
         N45\% & 0.45 & 2 & \rv{3.14} & \rv{30} & PCI & \rv{shear gradient}\\
         E65\% & 0.65 & 1 & \rv{6.28} & \rv{1-20} & ff-DDM & \rv{vorticity}\\
         E70\% & 0.70 & 1 & \rv{6.28} & \rv{1-20} & ff-DDM & \rv{vorticity}\\
         E74\% & 0.74 & 1 & \rv{6.28} & \rv{1-20} & ff-DDM & \rv{vorticity}\\
         E88\% & 0.88 & 1 & \rv{6.28} & \rv{1-20} & ff-DDM & \rv{vorticity}\\

    \end{tabular}
    \caption{Main features of the samples used in this study. M, N, and E refer to microgels, nanoparticles and emulsions, respectively. $\varphi$ designates the volume fraction for N and E, and the effective volume fraction for M, as defined in~\cite{philippe_glass_2018}. $T$ is the period of the oscillatory strain. PCI, SALS, and ff-DDM are photon correlation imaging, small angle light scattering and far field differential dynamic microscopy, respectively. \rv{The last column shows the orientation of the largest component of $\mathbf{q}$ with respect to the shear field, see} \textit{Experimental setups} for details.}
    \label{tab:samples}
\end{table}

\subsection*{Experimental setups} With the only exception of E samples, all experiments are performed with a home-made shear cell equipped with sliding parallel plates~\cite{aime_stress_controlled_2016}, sketched in~\SI~Fig. SI2. One plate is driven by a piezoelectric strain actuator (P602, from Physik Instrumente) and a force sensor (LC601, from Omega Engineering) measures the force applied by the actuator, enabling strain-controlled rheology experiments. The sample has a cross-sectional area of about $4~\mathrm{cm}^2$ and a thickness between $300$ and $700~\um$. For most experiments, shear rheology was coupled to a spatially-resolved Photon Correlation Imaging (PCI) apparatus~\rv{\cite{cipelletti_scattering_2016}}, collecting light scattered in a direction orthogonal to the shear and forming a scattering angle $\theta$ with the incoming beam. For samples N, we choose $\theta=141^\circ$, yielding a scattering vector $q=4\pi n_r \lambda^{-1} \sin(\theta/2)=30\um^{-1}$, where $n_r=1.33$ is the solvent refractive index and $\lambda=532.5$~nm the wavelength of laser light. In this case, $q$ is predominantly oriented along the shear gradient direction, with a minor component $q_x=3\um^{-1}$ along the vorticity direction. For \rv{M2s}, we choose $\theta=20^\circ$ such that $q=5\um^{-1}$ is predominantly oriented along the vorticity direction, with a minor component $q_z=0.9\um^{-1}$ along the shear gradient. For \rv{M40s}, experiments are performed using a different scattering geometry: a far-field small angle light scattering apparatus (SALS) enabling multiple scattering vectors to be probed simultaneously, oriented along \rv{both} the velocity and the vorticity direction, with $0.1~\um^{-1}\leq q \leq 5~\um^{-1}$ %\cite{tamborini_multiangle_2012}
. Data shown in the main text correspond to $q=4.8~\um^{-1}$, oriented along the vorticity direction. See~\SI~Figs SI1-\rv{SI3 for the schemes of the setups}.

For samples E, a different home-made shear cell with parallel, counter-translating plates displaced by a piezoelectric actuator is coupled to an inverted microscope with differential interference contrast (DIC) optics~\cite{knowlton_microscopic_2014} \rv{, see~\SI~Fig. SI3}. The acquired imaged region has a depth of field of $0.5~\um$, much smaller than the droplet size, and the imaged plane corresponds to the stagnation plane of the shear deformation. We analyze microscopy videos using far-field Differential Dynamic Microscopy (ff-DDM)~\cite{aime_probing_2019}, which yields an intensity correlation function equivalent to DLS, with scattering vectors $1~\um^{-1}\leq q \leq 20~\um^{-1}$ in the velocity-vorticity plane. Data presented in the main text correspond to $q=9~\um^{-1}$ along the vorticity direction. Data for more scattering vectors are included in~\SI~Fig. \rv{SI6-SI9}.

\subsection*{Characterization of the microscopic dynamics}

We quantify the microscopic dynamics via the two-time intensity correlation $g_2(t,\tau)-1 = \beta \langle I_p(t)I_p(t+\tau)\rangle/[\langle I_p(t)\rangle\langle I_p(t+\tau)\rangle]$, with $\beta$ a constant such that $g_2(t,\tau)-1 \rightarrow 1$ for $\tau \rightarrow 0$. $I_p$ is the scattered intensity collected by the $p-th$ pixel of the detector (for the PCI and SALS setups), or a component of the Fourier transform of the microscope images for sample E. $\langle \dots \rangle$ indicates the average over a set of pixels corresponding nearly to the same scattering vector or Fourier component.

In the stationary regime, we average $g_2-1$ over time $t$ to reduce noise before fitting Eq.~\ref{eqn:g1_dblexp} to the data. In Fig.~\ref{fig:phasediagram}a, the abscissa of the state points belonging to the coexistence region is calculated as the weighted average of the slow and fast relaxation rates, normalized by the relaxation rate at the critical point: $\rv{\Gamma_r = [\chi\Gamma_s + (\chi-1)\Gamma_f]/\Gamma_c}$. The slow and fast relaxation rates  are obtained by fitting Eq.~\ref{eqn:g1_dblexp} to the experimental $g_2-1$.

To quantify the temporal fluctuations of the dynamics, we inspect the $t$-dependence of the two-time correlation functions, with no $t$ averaging performed on $g_2(t,\tau)-1$. For each $t$, a relaxation time $\tau_r(t)$ is obtained from $\tau_r = \int_0 ^ \infty [g_2(t,\tau)-1]\mathrm{d}\tau$, a procedure more robust than fitting to Eq.~\ref{eqn:g1_dblexp} when dealing with correlation functions that are not averaged over $t$ and are thus more noisy. In Fig.~\ref{fig:order-transition}d we show $\sigma^2_{\tau_r}$, the temporal variance of $\tau_r(t)$.

We locate the range of the dynamic transition by determining the two strain amplitudes $\gamma_{0,s}$ and $\gamma_{0,e}$ such that $0.05 \le \chi \le 0.95$ for $\gamma_{0,s} \le \gamma_{0} \le \gamma_{0,e}$. The gray shaded regions in Figs.~\ref{fig:order-transition}d-f highlight the strain range $\gamma_{0,s} \le \gamma_{0} \le \gamma_{0,e}$. The normalized width of the transition shown in Fig.~\ref{fig:order-transition}a is defined as $w = (\gamma_{0,e}-\gamma_{0,s})/(\gamma_{0,e}+\gamma_{0,s})$.

\subsection*{Numerical solution of the model with disorder}
To study the effect of disorder on yielding, we implement our model (Eq.~\ref{eq:gamma_sh}) on a $D$-dimensional cubic lattice with periodic boundary conditions. Each site, $i$, is assigned a local relaxation rate, $\Gamma_i$, and each pair of neighbor sites is attributed a coupling constant, $\alpha_{ij}$, randomly drawn from a probability distribution, $P(\alpha)$. For a given strain amplitude $\gamma_0$ and starting from an initial configuration of local rates $\Gamma_i^{(init)}$, %for which in general Eq.~\ref{eq:gamma_sh} is not satisfied,
we seek a configuration of local rates $\Gamma_i^{(sol)}$ that satisfies Eq.~\ref{eq:gamma_sh} for all sites.
In our implementation, $\Gamma_i^{(sol)}$ is approached iteratively: at each step, a set of target site rates $\Gamma_i^{(t)}$ is computed through Eq.~\ref{eq:gamma_sh} using the current site rates $\Gamma_i^{(c)}$. $\Gamma_i^{(t)}$ then replaces $\Gamma_i^{(c)}$ for the following iteration, yielding a new set of target site rates. The convergence criterion is expressed in terms of a loss function $\mathcal{L}=\sum_i\left[\left(\log\Gamma_i^{(c)}-\log\Gamma_i^{(t)}\right)/\log\Gamma_i^{(t)}\right]^2$, which tends to 0 as $\Gamma_i^{(c)}$ approaches $\Gamma_i^{(sol)}$.%, as shown in the \SI.

Results shown in this paper are obtained for $D=2$ and a Log-Normal distribution of the coupling constants: $P(\alpha; \mu, \sigma) = (\alpha \sigma \sqrt{2\pi})^{-1} \exp[-(\ln \alpha-\mu)^2/(2 \sigma^2)]$, where the average $\bar\alpha$ and variance $\sigma_\alpha^2$ of the coupling constants are related to the parameters $\mu, \sigma$ of the Log-Normal distribution by $\bar\alpha = \exp(\mu + \sigma^2/2)$ and $\sigma_\alpha^2 = \bar\alpha^2(\exp\sigma^2-1)$. We quantify disorder by the dimensionless parameter $d = \sigma_\alpha^2/ \bar\alpha^2$. Representative results for different choices of $P(\alpha)$ are shown in~\SI~Fig.~\rv{SI13}, and exhibit no qualitative differences.

To mimic the effect of preshear in experiments, the iterative solution of the numerical model is typically initiated with a uniform distribution of local rates $\Gamma_i^{(init)}=\Gamma^{(0)}$. The effect of hysteresis shown in Figs.~\ref{fig:order-transition}b,c is studied by simulating an amplitude sweep experiment: $\gamma_0$ is first increased from low to high amplitudes and then decreased from high to low amplitudes, every time initiating the iterative calculation from the model solution for the previous amplitude.

%Log-Normal with mean $\alpha/z$, $z$ counting the number of nearest neighbors, and variance $\sigma_\alpha^2$.

\section*{Data availability}
The data that support the plots within this paper and other findings of this study are available from the corresponding authors upon request.

\section*{Code availability}
The code that support the plots within this paper and other findings of this study are available from the corresponding authors upon request.

%\nocite{*}

\bibliography{soft_glass}

%\begin{enumerate}
%    \item \perSA{Scheme of each setups}
%    \item \perSA{Info on $\gamma_0$ for Figs. 1a-c; info on model parameters for Fig. 1d-f.}
%    \item Fig. 4b: ``For the two points with the largest disorder (black symbols), simulations are not shown in panel (a) for the sake of clarity'' --> \perSA{bisognera' far vedere in~\SI~questi dati...}.
%    \item Fig. 4a: explain that in the coexistence region, the abscissa of experimental and numerical data is the average relaxation rate, define it.
%\end{enumerate}

%For the two points with the largest disorder (black symbols), simulations are not shown in panel (a) for the sake of clarity \perSA{bisognera' far vedere in~\SI~questi dati...}.

%\LC{Add a conclusive paragraph on perspectives, extension to other systems and/or driving mode, rationalization of results that were apparently conflicting (by highlighting different behaviors depending on the distance from a critical point, the model allows for rationalizing previous, apparently conflicting findings) etc.}

%\LC{emphasize somewhere that the dynamical coupling is the key ingredient that makes our stuff different from previous models. Another key point: the existence of a spontaneous rel rate in solid-like samples...Finally, can our model rationalize some previous observations? Which ones?}

%\LC{Note: at the end of the paper, we can mention interesting predictions of the catastrophe scenario: hysteresis, difference between glassy and supercooled or fluid samples (if not already discussed when first presenting current Fig. 3d). Other stuff?}

\begin{acknowledgments}
We thank E. D. Knowlton for help with the experiments on emulsions and L. Berthier for illuminating discussions. This work was funded by the French CNES,  ANR (grants No. ANR-14-CE32-0005, FAPRES, and ANR-20-CE06-0028, MultiNet), and by the EU (Marie Sklodowska-Curie ITN Supolen Grant 607937). LC acknowledges support from the Institut Universitaire de France.
\end{acknowledgments}

\section*{\label{sec:contributions}Author contributions}
SA, LR, DJP, and LC designed experiments. SA performed experiments and numerical simulations. SA and DT conceived the model. All authors analyzed the results, discussed and improved the model, and contributed to writing the paper.

\section*{Competing interests}
The authors declare no competing interests.

\section*{Additional information}
\textbf{Supplementary information} The online version contains supplementary material
available at [url to be inserted].

\textbf{Correspondence} and requests for materials should be addressed to SA or LC.

%%--FFFFFFFFFFFFFFFFFFFFFFFFFFFFFFFFFFFFFFFFFF
%\begin{figure}[ht]
%\includegraphics[width=8 cm,page=1,angle=90]{VdW-AimeTruzzolillo_dictionary}
%\includegraphics[width=8 cm,page=2,angle=90]{VdW-AimeTruzzolillo_dictionary}
%\includegraphics[width=8 cm,page=3,angle=90]{VdW-AimeTruzzolillo_dictionary}
%\caption{Correspondence between VdW EOS and Aime-Truzzolillo.
%\label{fig:vdW-AimeTruzzolillo}}
%\end{figure}
%--FFFFFFFFFFFFFFFFFFFFFFFFFFFFFFFFFFFFFFFFFFFFFFFF

%\appendix

\newpage

%\begin{figure}[ht]
\includegraphics[width=1\textwidth,page=1,angle=0]{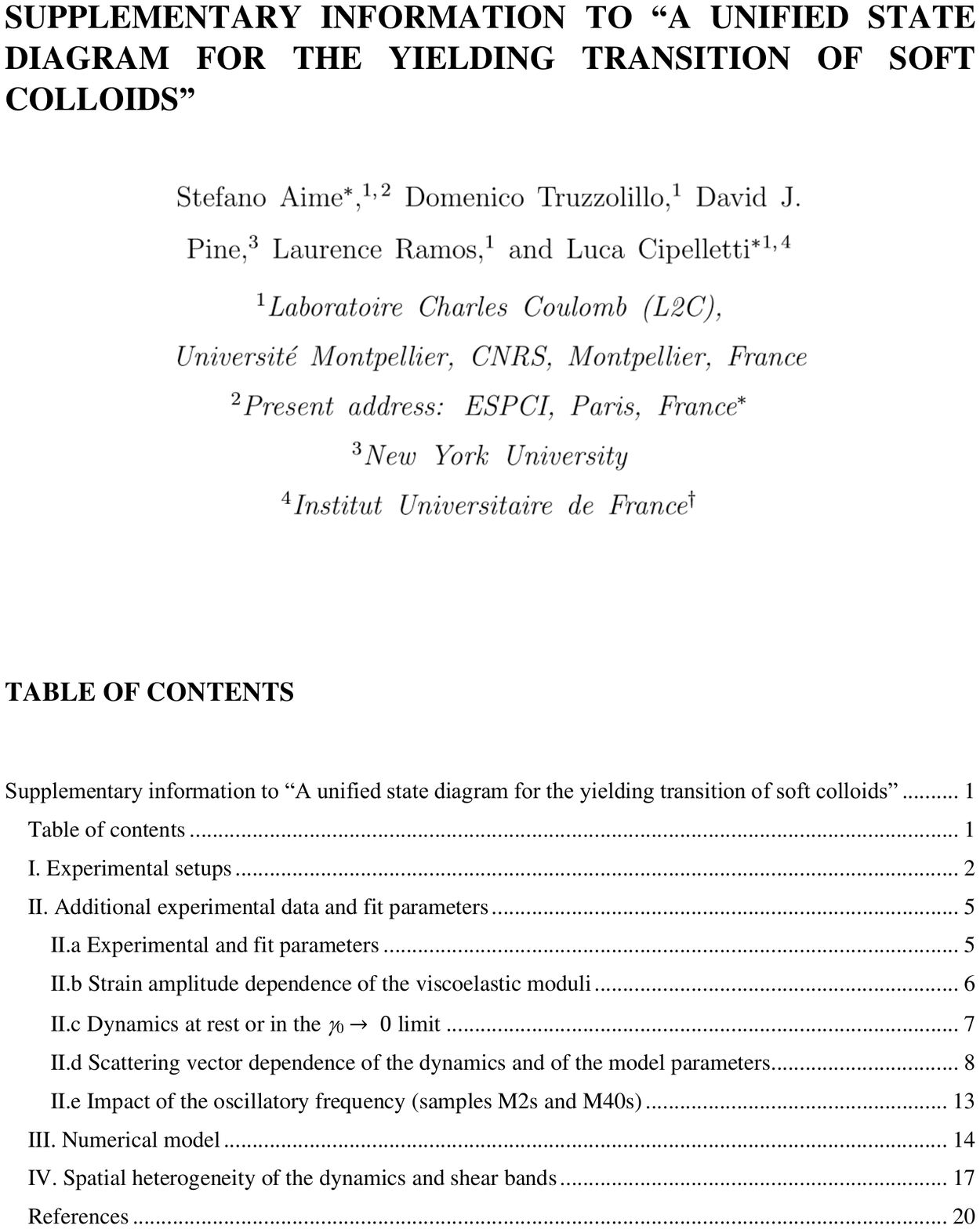}\\
\includegraphics[width=1\textwidth,page=2,angle=0]{221216_SI_black_version_nonumbers}\\
\includegraphics[width=1\textwidth,page=3,angle=0]{221216_SI_black_version_nonumbers}\\
\includegraphics[width=1\textwidth,page=4,angle=0]{221216_SI_black_version_nonumbers}\\
\includegraphics[width=1\textwidth,page=5,angle=0]{221216_SI_black_version_nonumbers}\\
\includegraphics[width=1\textwidth,page=6,angle=0]{221216_SI_black_version_nonumbers}\\
\includegraphics[width=1\textwidth,page=7,angle=0]{221216_SI_black_version_nonumbers}\\
\includegraphics[width=1\textwidth,page=8,angle=0]{221216_SI_black_version_nonumbers}\\
\includegraphics[width=1\textwidth,page=9,angle=0]{221216_SI_black_version_nonumbers}\\
\includegraphics[width=1\textwidth,page=10,angle=0]{221216_SI_black_version_nonumbers}\\
\includegraphics[width=1\textwidth,page=11,angle=0]{221216_SI_black_version_nonumbers}\\
\includegraphics[width=1\textwidth,page=12,angle=0]{221216_SI_black_version_nonumbers}\\
\includegraphics[width=1\textwidth,page=13,angle=0]{221216_SI_black_version_nonumbers}\\
\includegraphics[width=1\textwidth,page=14,angle=0]{221216_SI_black_version_nonumbers}\\
\includegraphics[width=1\textwidth,page=15,angle=0]{221216_SI_black_version_nonumbers}\\
\includegraphics[width=1\textwidth,page=16,angle=0]{221216_SI_black_version_nonumbers}\\
\includegraphics[width=1\textwidth,page=17,angle=0]{221216_SI_black_version_nonumbers}\\
\includegraphics[width=1\textwidth,page=18,angle=0]{221216_SI_black_version_nonumbers}\\
\includegraphics[width=1\textwidth,page=19,angle=0]{221216_SI_black_version_nonumbers}\\
\includegraphics[width=1\textwidth,page=20,angle=0]{221216_SI_black_version_nonumbers}\\
%\includegraphics[width=8 cm,page=2,angle=90]{VdW-AimeTruzzolillo_dictionary}
%\includegraphics[width=8 cm,page=3,angle=90]{VdW-AimeTruzzolillo_dictionary}
%\caption{Correspondence between VdW EOS and Aime-Truzzolillo.
%\label{fig:vdW-AimeTruzzolillo}}
%\end{figure}

%\includepdf[pages=-,pagecommand={},width=21.5cm,offset=0 -65 ]{221216_SI_black_version.pdf}

\end{document}